\documentclass[12pt]{iopart}
\usepackage{iopams}  
\usepackage{graphics}
\usepackage{epsfig}
\begin{document}

\title[Polaron Energy Spectrum in Quantum Dots]{Polaron Energy Spectrum in Quantum Dots}

\author{O.Z. Alekperov and N.M. Guseinov\footnote[3]{To whom correspondence should be addressed (semic@lan.ab.az)}}

\address{Institute of Physics, Azerbaijan National Academy of
Sciences, Javid av. 33, 370143, Baku, Azerbaijan}

\begin{abstract}

Energy spectrum of a weak coupling polaron  is considered in a cylindrical
quantum dot. An analytical expression for the polaron energy shift is
obtained using a modified pertubation theory.
\end{abstract}


\submitto {\JPCM}


\section{Introduction}

Investigation of electron spectrum in quantum dots has been attracted much
attention in the last decade as it has been technologically possible to
produce well characterized quasi-zero-dimensional structures. The energy
spectrum depens on the physical papameters of the quantum dot's material,
its shape and size. Also, the important factor influenses on the electron
spectrum in polar semiconductor is interaction between electrons and
longitudinal optical LO phonons. In low-dimesional structures the importance
of polaron effects increases. So, the knowledge of the polaron energy spectrum is important for understanding the infrared properties of quantum dots. One of the interesting experimental facts is the absence of the LO phonon bottleneck effect of photoexcited electrons. There is a number of papers proposed various reasons why the expected
bottleneck effect may be bypassed [1-3]. In particular, in the works [4,5]
the energy relaxation of the excited electrons in quantum dots was discussed
in connection with polaron effects. So, it is interesting to investigate
polaron spectrum in quantum dots in the region where the distance between
size-quantized leves equals to the LO-phonon energy.

In magneto-optical experiments in bulk crystals the electron-LO-phonon
interaction leads to anticrossing of the energy levels. Larsen was the first
to point out the level repulsion at $\omega _{c}=\omega _{LO}$, where $\omega _{c}$ is the cyclotron frequency and $\omega _{LO}$ is the frequency
of LO- phonon [6].The anticrossing of energy levels was observed in
absorption spectrum of quantum dots $InAs/GaAs$ in magnetic field in the
work [7]. Appearance of the anticrossing in the dependence of polaron levels
on the quantum dot radius theoretically was derived on the base of
two-levels system [5].

However, in order to obtain more correct expressions for the polaron energy
shift it is necessary to take into account the influence of the all
size-quantized levels. In present paper we consider an optical weak-coupling
polaron at low temperature in a cylindrical quantum dot. An analytical
expression for the polaron energy shift is obtained using a modified
perturbation theory.

\section{Polaron energy}

In the present paper we consider a disc shape quantum dot, i.e. a
cylindrical quantum dot with the radius essentially exceeding the  height. The
same situation usually is realized in experiments. Usually the diameter of
the disc exceeds its heigt by the order. Besides, we use the oscillator
model of the potential confining electron's movement along the cylindrical
axis $z$ with the frequency $\omega _{z}$ and in the plane of the disk with
the frequency $\omega $. According to the cosidered shape of the disc it is
assumed that $\omega _{z}\gg \omega $. It is considered that the levels
connected with confinement along $z$ are situated too over the ground state
and their influence can be neglected. It is suggested that the coupling
constant of a polar crystal in a quantum dot is too little (for $GaAs$ $\alpha \thickapprox 0.07$). In this connection we use the perturbation
theory for polaron energy shift. The suggested model of the quantum dot is
related to the following Schr\"{o}dinger equation for electron
noninteracting with phonons 

\begin{equation}
\fl \quad \left( \frac{\partial }{\partial \rho ^{2}}+\frac{1}{\rho }\frac{\partial }{\partial \rho }+\frac{1}{\rho ^{2}}\frac{\partial ^{2}}{\partial \varphi ^{2}}\right) \Psi +\frac{\partial ^{2}\Psi }{\partial z^{2}}+\left( \frac{2m_{0}E^{\left( 0\right) }}{\hbar ^{2}}-\frac{m_{0}^{2}\omega ^{2}\rho ^{2}}{\hbar ^{2}}-\frac{m_{0}^{2}\omega _{z}^{2}z^{2}}{\hbar ^{2}}\right) \Psi =0,
\end{equation}
where $m_{0}$ is electron's effective mass in the quantum dot, $\rho $, $%
\varphi $, $z$ are the cylindrical coordinates. The solution of the equation
(1) is well known

\begin{equation}
\Psi _{nm}=\frac{aa_{z}^{\frac{1}{2}}a_{nm}}{\pi ^{\frac{3}{4}}}\left( a\rho
\right) ^{\mid m\mid }L_{\frac{n-\left| m\right| }{2}}^{\left| m\right|
}\left( a^{2}\rho ^{2}\right) \exp \left( -\frac{a^{2}\rho ^{2}}{2}-\frac{%
a_{z}^{2}z^{2}}{2}+im\varphi \right) ,
\end{equation}
where $L_{n}^{\lambda }\left(x\right) $ are the associated Laguerre polynomials, $a^{2}=\frac{%
m_{0}\omega }{\hbar }$; $a_{z}^{2}=\frac{m_{0}\omega _{z}}{\hbar }$; $%
a_{nm}^{2}=\frac{\left( \frac{n-\left| m\right| }{2}\right) !}{\left( \frac{%
n+\left| m\right| }{2}\right) !}$ ($n=0,1,2,...$, $m=0,\pm 2,\pm 4,...,\pm n$
if $n$ is even, $m=\pm 1,\pm 3,...,\pm n$ if $n$ is odd and $m=0$ for $n=0$).
The energy spectrum is:

\begin{equation}
E_{n}^{\left( 0\right) }=\frac{\hbar \omega _{z}}{2}+\hbar \omega \left(
n+1\right) .
\end{equation}

The electrons are assumed to be coupled to dispersionless LO-phonons of the
bulk crystal. The potential yielded by one LO-phonon is

\begin{equation}
\varphi _{q}=\frac{\Lambda }{q}e^{iq_{\bot }\rho \cos \varphi +iq_{z}z},
\end{equation}
where $\Lambda =i\hbar \omega _{0}\left( 4\pi \alpha /\gamma _{0}V\right) $, 
$q$ is the phonon wavevector, $q_{\perp }$and $q_{z}$ are the components of
the phonon wavevector laying in the plane of the disc and along $z$ axes
correspondingly, $q=\sqrt{q_{\perp }^{2}+q_{z}^{2}}$, $\gamma _{0}^{2}=\frac{%
2m_{0}\omega _{0}}{\hbar }$, $\hbar \omega _{0}$ is the energy of LO phonon
at $q=0$, $V$ is the quantum dot's volume.

The matrix element corresponding to the emission of the LO phonon is

\begin{equation}
M_{nm\longrightarrow n^{\prime }m^{\prime }}=\int \Psi _{nm}e\varphi
_{q}^{*}\Psi _{n^{\prime }m^{\prime }}^{*}dV,
\end{equation}
where $e$ is the electron charge. Using the expressions (2), (4) and (5) we obtain

\begin{equation}
M_{nm\rightarrow n^{\prime }m^{\prime }}=\frac{2\Lambda
a_{nm}a_{n^{\prime }m^{\prime }}\left( -i\right) ^{m^{\prime }-m}}{q}e^{-%
\frac{q_{z}^{2}}{4a_{z}^{2}}}I_{nm\rightarrow n^{\prime }m^{\prime
}}\left( \frac{q_{\perp }}{2a}\right) ,
\end{equation}
where 
\begin{equation*}
I_{nm\longrightarrow n^{\prime }m^{\prime }}\left( y\right)
=\int_{0}^{\infty }x^{\left| m\right| +\left| m^{\prime }\right| +1}L_{\frac{%
n-\left| m\right| }{2}}^{\left| m\right| }\left( x^{2}\right) L_{\frac{%
n^{\prime }-\left| m^{\prime }\right| }{2}}^{\left| m^{\prime }\right|
}\left( x^{2}\right) e^{-x^{2}}J_{m^{\prime }-m}\left( 2xy\right) ,
\end{equation*}
$J_{m^{\prime }-m}\left( 2xy\right) $ is the Bessel function.

It is assumed that the difference between conductivity bands of the material
of quantum dot and surroundings is quite large and a sufficient number of
the quantum levels can exist inside the quantum well. Therefore, while
calculating the polaron energy, we shall take into account the electron's
transitions to the all levels. Far from the resonance $\left( E_n^{(0)}- E_{n'} ^{(0)}\right)= \hbar \omega_0$, the polaron correction to the energy in the second order of the perturbation theory is
defined as follows:

\begin{equation}
\Delta E_{n}=\sum\limits _{m,n^{\prime },m^{\prime }, \overrightarrow{q}} \frac{\left| M_{nm\rightarrow n^{\prime }m^{\prime }}\right| ^{2}}{%
E_{n}^{\left( 0\right) }-E_{n^{\prime }}^{\left( 0\right) }-\hbar \omega _{0}%
}.
\end{equation}

The expression (7) differs from the one, usually being
used for calculation of a correction to the ground state, by the additional
sum over degenerate states of the level $E_{n}^{\left( 0\right) }$
(summation over $m$). We shall substitute (3) and (6) in (7) and replace the
summation over $\overrightarrow{q}$ with integration. For simplicity we
shall consider the limit $\omega _{z},a_{z}\rightarrow \infty $ related
to the case of zero height cylinder. According to this limit the integral
over $q_{z}$ is equal to:

\begin{equation*}
\int_{-\infty }^{\infty }\frac{\exp \left( -\frac{q_{z}^{2}}{2a_{z}^{2}}%
\right) }{q_{\perp }^{2}+q_{z}^{2}}dq_{z}\mid _{q_{z}\rightarrow \infty
}=\frac{\pi }{q_{\perp }} .
\end{equation*}

As a result we shall obtain the following expression of the polaron energy
shift normalized by $\alpha \hbar \omega _{0}$:

\begin{equation}
\frac{\Delta E_{n}}{\alpha \hbar \omega _{0}}=-8\left( \frac{\Gamma }{2}%
\right) ^{\frac{1}{2}}\sum_{m,n^{\prime },m^{\prime }}\frac{%
a_{nm}^{2}a_{n^{\prime }m^{\prime }}^{2}J_{nm\longrightarrow n^{\prime
}m^{\prime }}}{\left( n^{\prime }-n\right) \Gamma +1}
\end{equation}

\[
J_{nm\longrightarrow n^{\prime }m^{\prime }}=\int_{0}^{\infty }\left|
I_{nm\longrightarrow n^{\prime }m^{\prime }}\right| ^{2}dy, 
\]
\begin{figure}[h!]
\epsfig{file=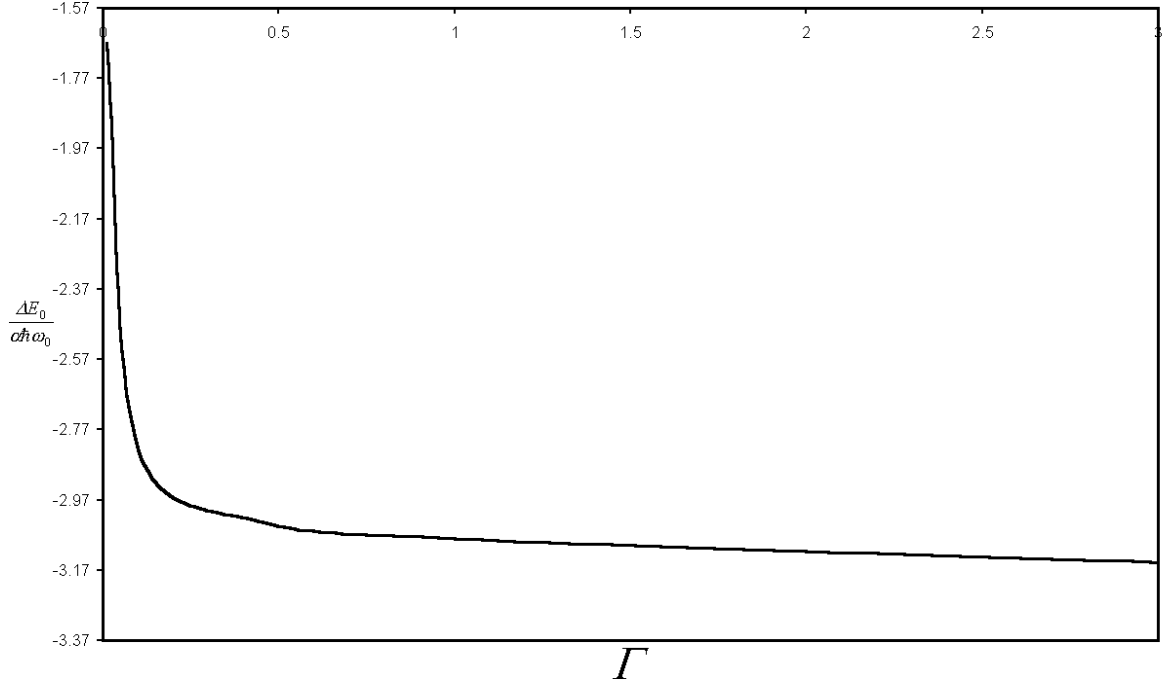,height=9.73cm,width=15.82cm}
\caption{Dependence of the polaron ground state shift $\Delta E_{0}/\alpha \hbar
\omega _{0}$ on $\Gamma $.}
\end{figure}
where $\Gamma =\omega /\omega _{0}$. As the parameter $\Gamma $ is proportional to $1/R$,
where $R$ is the quantum dot radius, the limit $\Gamma \rightarrow \infty $
corresponds to the ultraquantum limit. On the other hand, the limit $\Gamma
\rightarrow 0$ corresponds to the two-dimensional case of a plane. The
polaron energy shift of the ground state $\Delta E_{0}/\alpha \hbar \omega
_{0}$ calculated as a function of the parameter $\Gamma $ is plotted in
Fig.1. The quantity of $\Delta E_{0}/\alpha \hbar \omega _{0}$ trends to the
well known value $-\pi /2$ for two-dimensional case in the limit $\Gamma
\rightarrow 0.$ In the case of $\Gamma \rightarrow \infty $ the quantity of $%
\Delta E_{0}/\alpha \hbar \omega _{0}$ diverges proportionally to $\Gamma ^{%
\frac{1}{2}}$. It is known that the divergence is taken place also in the
case of a quantum wire if the wire radius tends to zero [8]. However, in
one-dimensional case there is more weak logarithmic divergence. Now we shall
calculate the polaron energy shift of the level $E_{1}^{\left( 0\right) }$
for any value of $\Gamma $ including the resonace region. As it is seen from
the expression (9) for $n=1$, the term corresponding to $n^{\prime }=0, $%
diverges in the limit $\Gamma \rightarrow 1$. The formula (9) is not
applicable near the resonance $\Gamma =1$. In order to calculate the energy
shift of the level $E_{1}^{\left( 0\right) }$ in this area we shall exclude
from (9) the term corresponding to the transition from the state $n=1$ to $%
n^{\prime }=0$ and shall take into account this contribution using the
perturbation theory applicable in the case of degeneracy of two levels $%
E_{1}^{\left( 0\right) }$ and $E_{0}^{\left( 0\right) }+\hbar \omega _{0}$.
We shall consider the system consisted from an electron, which has the state 
$n=0,m=0$ of the level $E_{0}^{\left( 0\right) }$ and two degenerate states $%
n=1,m=\pm 1$ of the level $E_{1}^{\left( 0\right) }$, and a phonon which
will have the occupation number $n_{q}=0$ if the electron is on the level $%
E_{0}^{\left( 0\right) }$ and $n_{q}=1$ if the electron is on the level $%
E_{1}^{\left( 0\right) }$. So, the united noninteracting electron-phonon
system may exist on the level $E_{1}^{\left( 0\right) }$ with two degenerate
states $\left| n=1,m=\pm 1,n_{q}=0\right\rangle $ and on the level $%
E_{0}^{\left( 0\right) }+\hbar \omega _{0}$ with the state $\left|
n=0,m=0,n_{q}=1\right\rangle $. Then including the Fr\"{o}hlich coupling one
can easy obtain the following expression for the energy of the two-level
system:

\begin{equation}
E=\frac{E_{1}^{\left( 0\right) }+E_{0}^{\left( 0\right) }+\hbar \omega _{0}}{%
2}\pm \sqrt{\frac{1}{4}\left( E_{1}^{\left( 0\right) }-E_{0}^{\left(
0\right) }-\hbar \omega _{0}\right) ^{2}+\left| V_{01}\right| ^{2}},
\end{equation}
where $E_{0}^{\left( 0\right) }=\hbar \omega $, $E_{1}^{\left( 0\right)
}=2\hbar \omega $, $\left| V_{01}\right| ^{2}=\sum\limits_{\overrightarrow{q}%
}\left( \left| M_{0,0\rightarrow 1,1}\right| ^{2}+\left| M_{0,0\rightarrow
1,-1}\right| ^{2}\right) $.

From (9) we can define $\Delta E_{1}=E-E_{1}^{\left( 0\right) }$ in the
following form:

\begin{equation}
\frac{\Delta E_{1}}{\alpha \hbar \omega _{0}}=\frac{1-\Gamma }{2\alpha }\pm 
\frac{1}{\alpha }\sqrt{\left( \frac{1-\Gamma }{2}\right) ^{2}+\frac{\left|
V_{01}\right| ^{2}}{\left( \hbar \omega _{0}\right) ^{2}}},
\end{equation}
where $\frac{\left| V_{01}\right| ^{2}}{\left( \hbar \omega _{0}\right) ^{2}}%
=16\left( \frac{\Gamma }{2}\right) ^{\frac{1}{2}}\alpha J_{1,1\rightarrow
0,0}$.

The sign plus corresponds to the region $\Gamma \eqslantgtr 1$, the sign
minus to the region $\Gamma \eqslantless 1$. The contribution into the
energy shift $\Delta \left( E_{0}+\hbar \omega _{0}\right) =E-E_{0}^{\left(
0\right) }-\hbar \omega _{0}$ corresponding to the same transition is
defined by the formula (11), but with the sign plus in the region $\Gamma
\eqslantless 1$ and with the sign minus in the region $\Gamma \eqslantgtr 1$.

\begin{equation}
\frac{\Delta \left( E_{0}+\hbar \omega _{0}\right) }{\alpha \hbar \omega _{0}%
}=-\frac{1-\Gamma }{2\alpha }\pm \frac{1}{\alpha }\sqrt{\left( \frac{%
1-\Gamma }{2}\right) ^{2}+\frac{\left| V_{01}\right| ^{2}}{\left( \hbar
\omega _{0}\right) ^{2}}}
\end{equation}

The value of the integral $J_{1,1\rightarrow 0,0.}$is $0.039$.

\begin{figure}[h!]
\epsfig{file=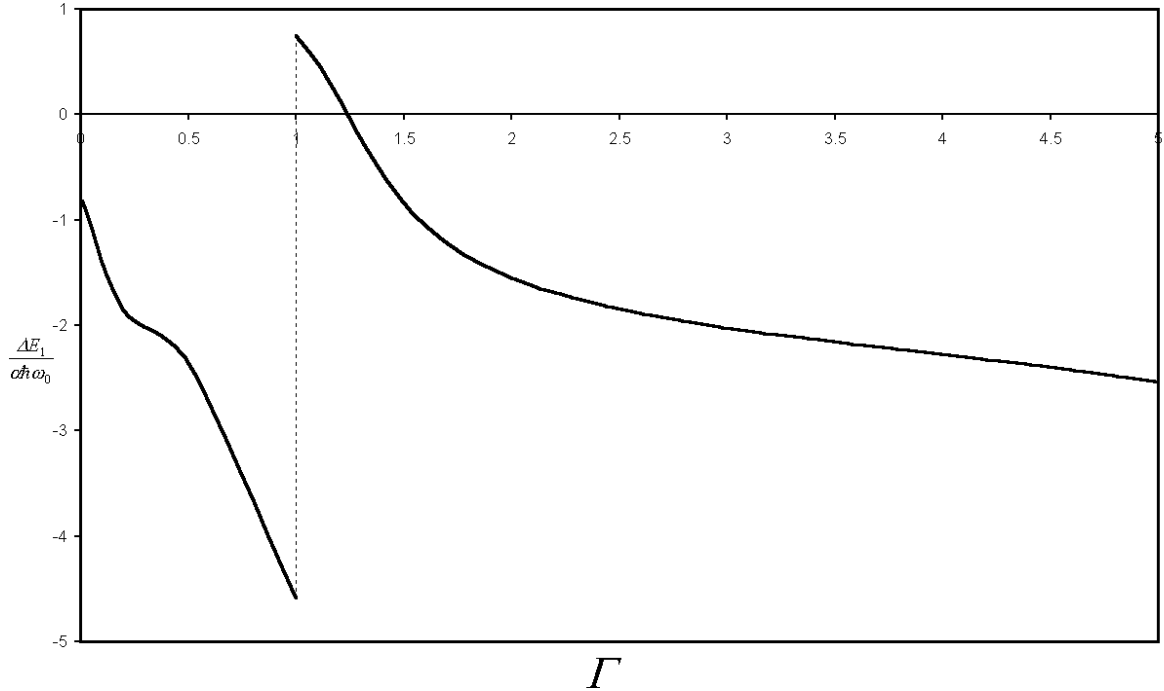,height=9.73cm,width=15.82cm}
\caption{Dependence of the polaron energy shift $\Delta E_{1}/\alpha \hbar
\omega _{0}$ on $\Gamma $.}
\end{figure}
\begin{figure}[h!]
\epsfig{file=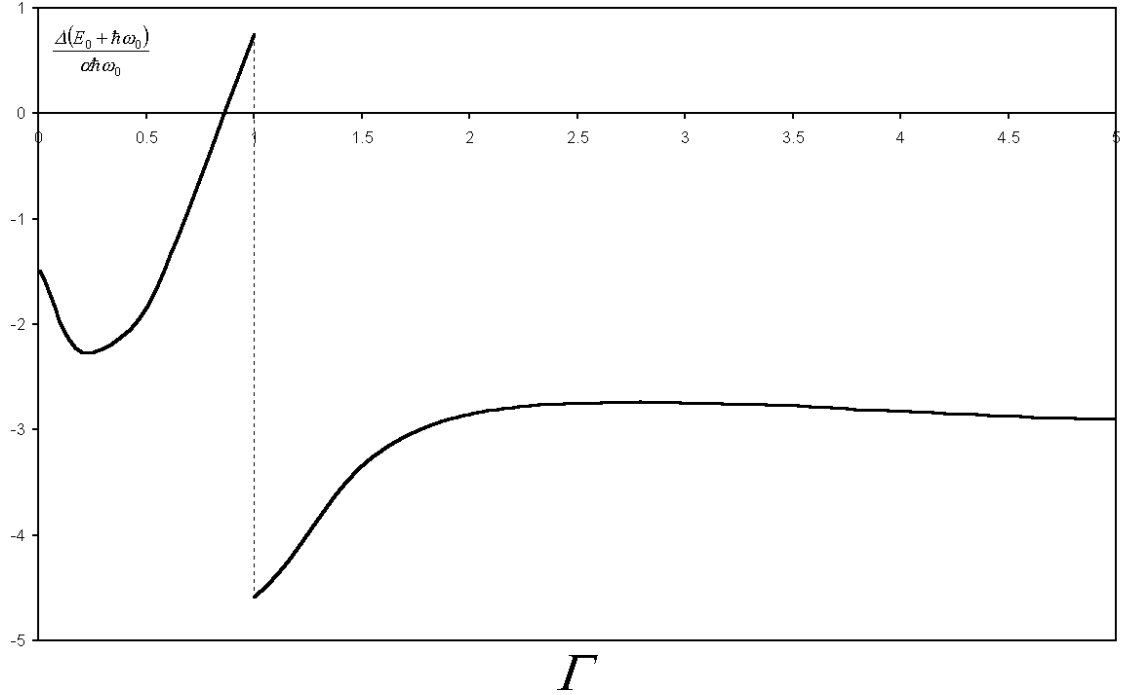,height=9.73cm,width=15.82cm}
\caption{Dependence of the $\Delta \left( E_{0}+\hbar \omega _{0}\right)
/\alpha \hbar \omega _{0}$ on $\Gamma $.}
\end{figure}
The total shift $\Delta E_{1}/\alpha \hbar \omega _{0}$ and $\Delta \left(
E_{0}+\hbar \omega _{0}\right) /\alpha \hbar \omega _{0}$ is calculated
excluding from (9) the term related to the transition $n=1,m=\pm
1\rightarrow n=0,m=0$ and adding the contribution defined by (11) and (12)
respectively. The shifts $\Delta E_{1}/\alpha \hbar \omega _{0}$ and $\Delta
\left( E_{0}+\hbar \omega _{0}\right) /\alpha \hbar \omega _{0}$ as the
functions of $\Gamma $ are plotted in fig.2 and fig.3 respectively for $%
\alpha \approx 0.07$. It is seen from fig.2 and fig.3 that $\Delta E_1 |_{\Gamma \rightarrow 1 \mp 0} = \Delta \left(E_0+\hbar \omega _0 \right) |_{\Gamma \rightarrow 1 \pm 0}$. The picture of the levels $E_{1}/\hbar \omega _{0}$
and $\left( E_{0}+\hbar \omega _{0}\right) /\hbar \omega _{0}$ is
represented in fig.4, where $E_{1}^{\left( 0\right) }/\hbar \omega
_{0}=2\Gamma $ and $\left( E_{0}^{\left( 0\right) }+\hbar \omega _{0}\right)
/\hbar \omega _{0}=\Gamma +1$ are represented by the dashed lines. In the
resonance area of the energy levels the corresponding wave function are represented
as a superposition of the states $\left| n=1,m=\pm 1,n_{q}=0\right\rangle $
and $\left| n=0,m=0,n_{q}=1\right\rangle $.
\begin{figure}[h!]
\epsfig{file=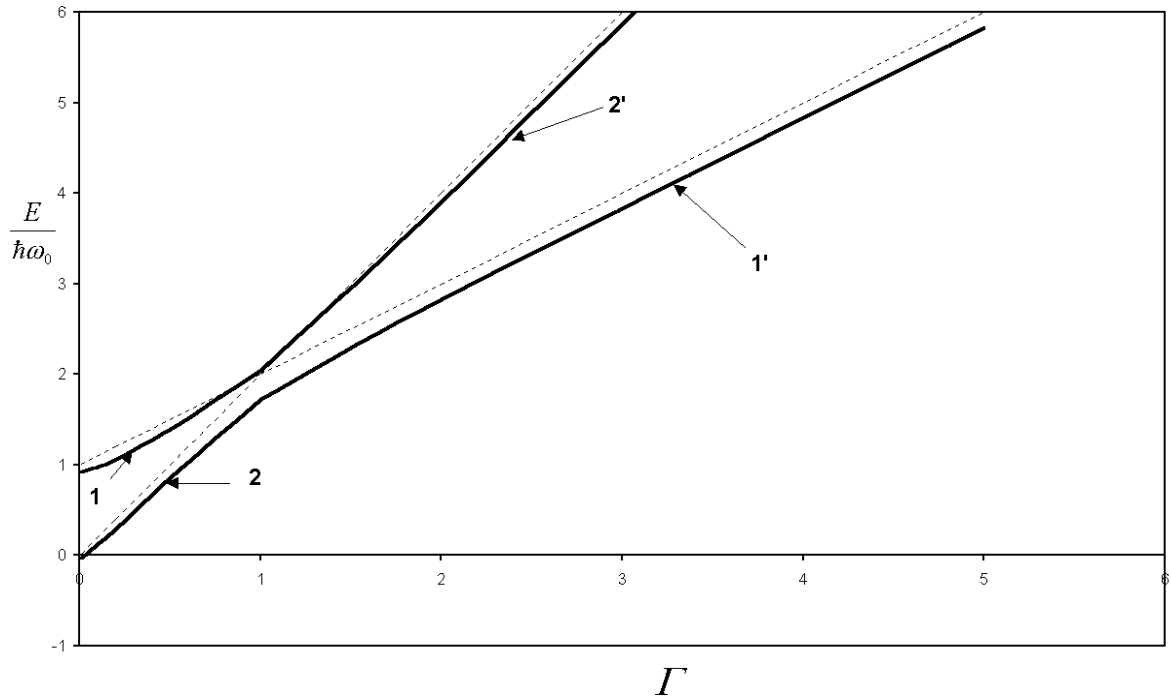,height=9.73cm,width=15.82cm}
\caption{Dependence of the levels $E_{1}/\hbar \omega _{0}$and $\left(
E_{0}+\hbar \omega _{0}\right) /\hbar \omega _{0}$on $\Gamma $. The levels $%
E_{1}^{\left( 0\right) }/\hbar \omega _{0}$ and $\left( E_{0}^{\left(
0\right) }+\hbar \omega _{0}\right) /\hbar \omega _{0}$ are plotted by
dashed lines. Here $1$ - $\left(E_0+\hbar \omega _0 \right) / \hbar \omega _0$ in $\Gamma \leq 1$ region; $1'$ - $\left(E_0+\hbar \omega _0 \right) / \hbar \omega _0$ in $\Gamma \geq 1$ region; $2$ - $E_1 / \hbar \omega _0$ in $\Gamma \leq 1$ region; $2'$ - $E_1 / \hbar \omega _0$ in $\Gamma \geq 1$ region.}
\end{figure}

\section{Conclusion}

Using the perturbation theory the polaron energy shift was obtained. The
polaron shift of ground state tends to the well known value $-\pi /2$ for
two-dimensional system in the case of $\Gamma \rightarrow 0.$In the limit of 
$\Gamma \rightarrow \infty $ the polaron shift diverges more sharp than in
the case of quantum wire. when its radius tends to zero.

Using the modified perturbation theory the anticrossing of the polaron
levels $E_{1}$and $E_{0}+\hbar \omega _{0}$ was obtained near the resonance
region.

The picture like the one plotted in fig.4 was obtained in experimental
dependence of polaron spectrum on the magnetic field directed along axes $z$
of the cylindrical quantum dots [7]. In the presence of magnetic field the
derived formulas are the same with substitution of $\omega $ by $\left(
\omega ^{2}+\omega _{c}^{2}/4\right) ^{\frac{1}{2}}$. The role of the
parameter $\Gamma $ will play the magnitude $\left( \omega ^{2}+\omega
_{c}^{2}/4\right) ^{\frac{1}{2}}/\omega _{0}$.

Note that the splitting of the levels $E_{1}$ and $E_{0}+\hbar \omega _{0}$
in light absorption experiment will be observed also in the absence of
magnetic field if $\omega $ $\thickapprox \omega _{0}$. Besides the
allowed transition $E_{0}\rightarrow $ $E_{1}$, the transition $%
E_{0}\rightarrow $ $E_{0}+\hbar \omega _{0}$ may also take place because
the wave functions of these levels are mixed in the anticrossing region. Far from the resonance the level $E_{0}+\hbar \omega _{0}$ has not the
particular physical meaning, however, the states corresponding to this level
in the resonance region may be realized in experiments on light absorption.

\end{document}